\documentclass{cupconf}
\usepackage{graphicx}

  \checkfont{eurm10}
  \iffontfound
    \IfFileExists{upmath.sty}
      {\typeout{^^JFound AMS Euler Roman fonts on the system,
                   using the 'upmath' package.^^J}%
       \usepackage{upmath}}
      {\typeout{^^JFound AMS Euler Roman fonts on the system, but you
                   dont seem to have the}%
       \typeout{'upmath' package installed. cupconf.cls can take advantage
                 of these fonts,^^Jif you use 'upmath' package.^^J}%
      }
  \else
  \fi

  \checkfont{msam10}
  \iffontfound
    \IfFileExists{amssymb.sty}
      {\typeout{^^JFound AMS Symbol fonts on the system, using the
                'amssymb' package.^^J}%
       \usepackage{amssymb}%
         \let\leq=\leqslant
         \let\geq=\geqslant
      }{}
  \fi

  \IfFileExists{amsbsy.sty}
    {\typeout{^^JFound the 'amsbsy' package on the system, using it.^^J}%
     \usepackage{amsbsy}}
    {\providecommand\boldsymbol[1]{\mbox{\boldmath $##1$}}}

\newsavebox{\astrutbox}
\sbox{\astrutbox}{\rule[-5pt]{0pt}{20pt}}

\def\um     {$\mu$m}

\def\nwm2sr {\wisk{\rm nW/m^2/sr\ }}
\def\nw2m4sr {\wisk{\rm nW^2/m^4/sr\ }}
\title[Cosmic infrared background and early stars]{Cosmic infrared background and Population III: an overview}

\author[A. Kashlinsky]{A. Kashlinsky}

\affiliation{Code 665, Observational Cosmology Lab, Goddard Space
Flight Center, Greenbelt MD 20771\\
e-mail: kashlinsky@milkyway.gsfc.nasa.gov}

\date{?? and in revised form ??}
\begin{document}

\maketitle

\begin{abstract}
We review the recent measurements on the cosmic infrared
background (CIB) and their implications for the physics of the
first stars era, including Population III. The recently obtained
CIB results range from the direct measurements of CIB fluctuations
from distant sources using deep Spitzer data to strong upper
limits on the near-IR CIB from blazar spectra. This allows to
compare the Population III models with the CIB data to gain direct
insight into the era of the first stars and the formation and
evolution of Population III and the microphysics of the feedback
processes in the first halos of collapsing material. We also
discuss the cosmological confusion resulting from these CIB
sources and the prospects for resolving them individually with
NASA's upcoming space instruments such as the {\it JWST}.
\end{abstract}

\firstsection 
\section{Introduction}

The very first stars to form in the Universe, commonly called
Population III, are now thought to have been very massive stars
forming out of primordial metal-free gas at redshifts exceeding
$z\sim 10$ (see recent review by Bromm \& Larson 2004). Assuming
that the density field responsible for structure formation is
given by the $\Lambda$CDM model, the first collapsing haloes
hosting such stars may be too faint to be observed with the
present telescopes. Their studies may, however, be possible via
the cumulative radiation emitted by the first luminous objects,
most of which by now has been shifted into the near-IR wavelengths
of $\sim 1-10$ \um. We review here the connection between the
massive Population III stars and the cosmic infrared background
measurements and the implications of the latter for the duration,
nature and abundance of luminous objects that comprised the era of
the first stars.

The cosmic infrared background (CIB) is a repository of emissions
throughout the entire history of the Universe. Cosmic expansion
shifts photons emitted in the visible/UV bands at high $z$ into
the near-IR (NIR) and the high-$z$ NIR photons appear today in
mid- to far-IR. Consequently, the NIR part of the CIB spectrum
(1$\mu$m$ < \lambda < $10 $\mu$m) probes the history of direct
stellar emissions from the early Universe, and the longer
wavelengths contain information about the early dust production
and evolution. The recent years have seen significant progress in
CIB studies, both in identifying and/or constraining its mean
level (isotropic component) and fluctuations (see Kashlinsky 2005a
for a recent review).

The CIB contains emissions also from objects inaccessible to
current (or even future) telescopic studies and can, therefore,
provide unique information on the history of the Universe at very
early times. One particularly important example of such objects
are Population III stars (hereafter Pop~III), the still elusive
zero-metallicity stars expected to have preceded normal stellar
populations seen in the farthest galaxies to-date. Throughout this
review we will use the term "era of the first stars", or "Pop~III
era", with the understanding that the actual era may be composed
of objects of various nature from purely zero-metallicity stars,
to low- metallicity stars to even possibly mini-quasars whose
contribution to the CIB is driven by energy released by
gravitational accretion, as opposed to stellar nucleosynthesis.

Pop~III stars are thought to have preceded the normal
metal-enriched stellar populations, but even if massive and
luminous they are inaccessible to direct observations by current
telescopes because of the high redshits at which they are located.
Extensive numerical investigations, as discussed by Bromm and
Norman in this meeting, suggest that they had to be very massive,
forming out of density fluctuations specified by the standard
$\Lambda$CDM model. If massive, they are expected to have left a
significant level of diffuse radiation shifted today into the IR,
and it was suggested that the CIB contains a significant
contribution from Pop~III in the near-IR, manifest in both its
mean level and its anisotropies (e.g. Santos et al 2002,
Salvaterra \& Ferrara 2003, Cooray et al 2004, Kashlinsky et al
2004; see also review by Kashlinsky 2005a). This notion has
recently received strong support from measurements of CIB
anisotropies in deep Spitzer/IRAC images (Kashlinsky et al 2005)
as will be discussed below.

The structure of this review is as follows: Sec. \ref{sec:cib_dc}
summarizes the current measurements of the mean levels of the CIB
and the contributions to them from the observed galaxy
populations. Sec. \ref{sec:pop3-theory} discusses the theoretical
connection between the CIB and the first (massive) stars and Sec.
\ref{sec:gamma-rays} gives the limits on the near-IR CIB from
measurements of absorption in the high-energy spectra of
cosmological sources. Sec. \ref{sec:spitzer_kamm} reviews the
recent Spitzer-based measurements of CIB fluctuations from early
epochs and their interpretation in terms of the nature and the
epochs of the contributing sources. Finally, the prospects for
resolving these sources in future measurements are summarized in
Sec. \ref{sec:confusion} followed with conclusions in Sec.
\ref{sec:conclusions}.

The AB magnitude system is adopted throughout; the conversion to
fluxes is simple as zero AB magnitude corresponds to the flux of
3631 Jy.

\section{Direct CIB measurements}
\label{sec:cib_dc}

Galactic and Solar System foregrounds are the major obstacles to
space-based CIB measurements. Galactic stars are the main
contributors at near-IR ($<$ a few \um), zodiacal light from the
dust in the Solar system dominates between $\sim$10 and $\sim$50
\um, and Galactic cirrus emission produces most of the foreground
at IR wavelengths at $> 50$ \um. Accurately removing the
foregrounds presents a challenge and many techniques have been
developed to do this as well as possible. Stars can be removed in
surveys with fine angular resolution or by statistically
extrapolating the various stellar contributions to zero. Zodiacal
light contributions usually are removed using the DIRBE zodi model
or its derivatives. Galactic cirrus and zodiacal light are both
intrinsically diffuse, but are fairly homogeneous adding to the
effectiveness of CIB fluctuations studies at mid- to far-IR
(Kashlinsky et al 1996a,b).

\begin{figure}
 \includegraphics[height=2.5in,width=5.5in]{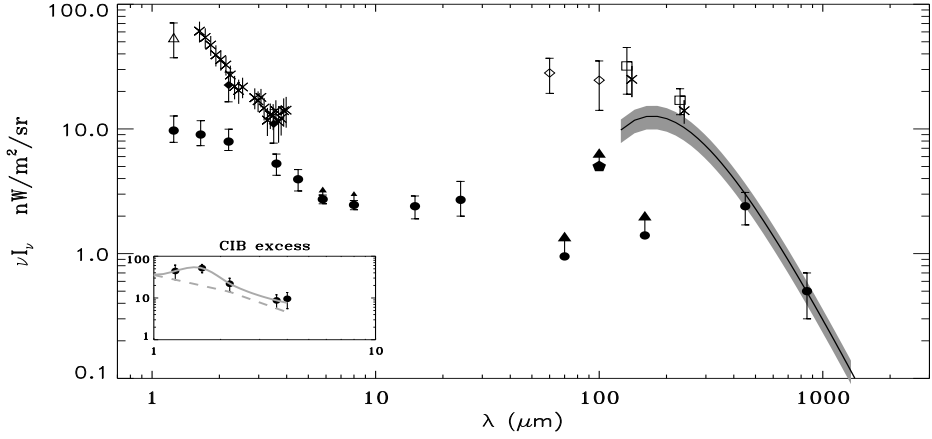}
 \caption{ CIB
 and ``ordinary" galaxy (OG) contributions
(filled circles) vs wavelength. The numbers are adopted from Fig.
9 of Kashlinsky (2005a and references therein) and are discussed
at length there. Briefly, the CIB fluxes from IRTS measurements
are shown with crosses from Matsumoto et al (2005), diamonds are
from Gorjian et al (2000) at 2.2 \um, and from Dwek \& Arendt
(1998) and Wright \& Reese (2000) at 3.5 \um. The flux from
ordinary galaxies is shown with filled circles and is taken from
HST counts out to 2.2 \um\ (Madau \& Pozzetti 2000) and from
Spitzer/IRAC counts at 3.6 and 4.5 \um\ (Fazio et al 2004). At
$\lambda>$10\um\ no CIB excess was observed and the levels of CIB
are consistent with the net contribution from OG. The inset shows
the NIRBE spectrum, $\nu I_\nu$ in nW/m$^2$/sr vs $\lambda$ in
\um, from Kashlinsky (2005a). The two thick light-shaded lines
show the ranges of the NIRBE spectrum suggested by the data: the
solid line goes through the central points, while the dashed line
grazes the lower edges of the data.} \label{fig:cib_dc}
\end{figure}

In the near-IR CIB detections are difficult because of the
substantial foreground by Galactic stars. Claims of detections of
the mean isotropic part of the CIB are based on various analyses
of DIRBE and IRTS data  (Dwek \& Arendt 1998, Matsumoto et al
2005, Wright \& Reese 2000, Gorjian et al 2001, Cambresy et al
2001). The measurements agree with each other, although the
methods of analysis and foreground removal differ substantially.
They also agree with the measured amplitude of CIB fluctuations
using DIRBE data (Kashlinsky \& Odenwald 2000). The results seem
to indicate fluxes significantly exceeding those from observed
galaxy populations. Fig. \ref{fig:cib_dc} summarizes the current
CIB measurements. Only the near-IR CIB at $\lambda < 10$ \um\ is
relevant for the discussion that follows as we will not discuss
here the possible dust contribution from the high $z$ contributing
to (re)emissions contributing to CIB at longer wavelengths.

\begin{figure}
 \includegraphics[height=3.5in,width=5.5in]{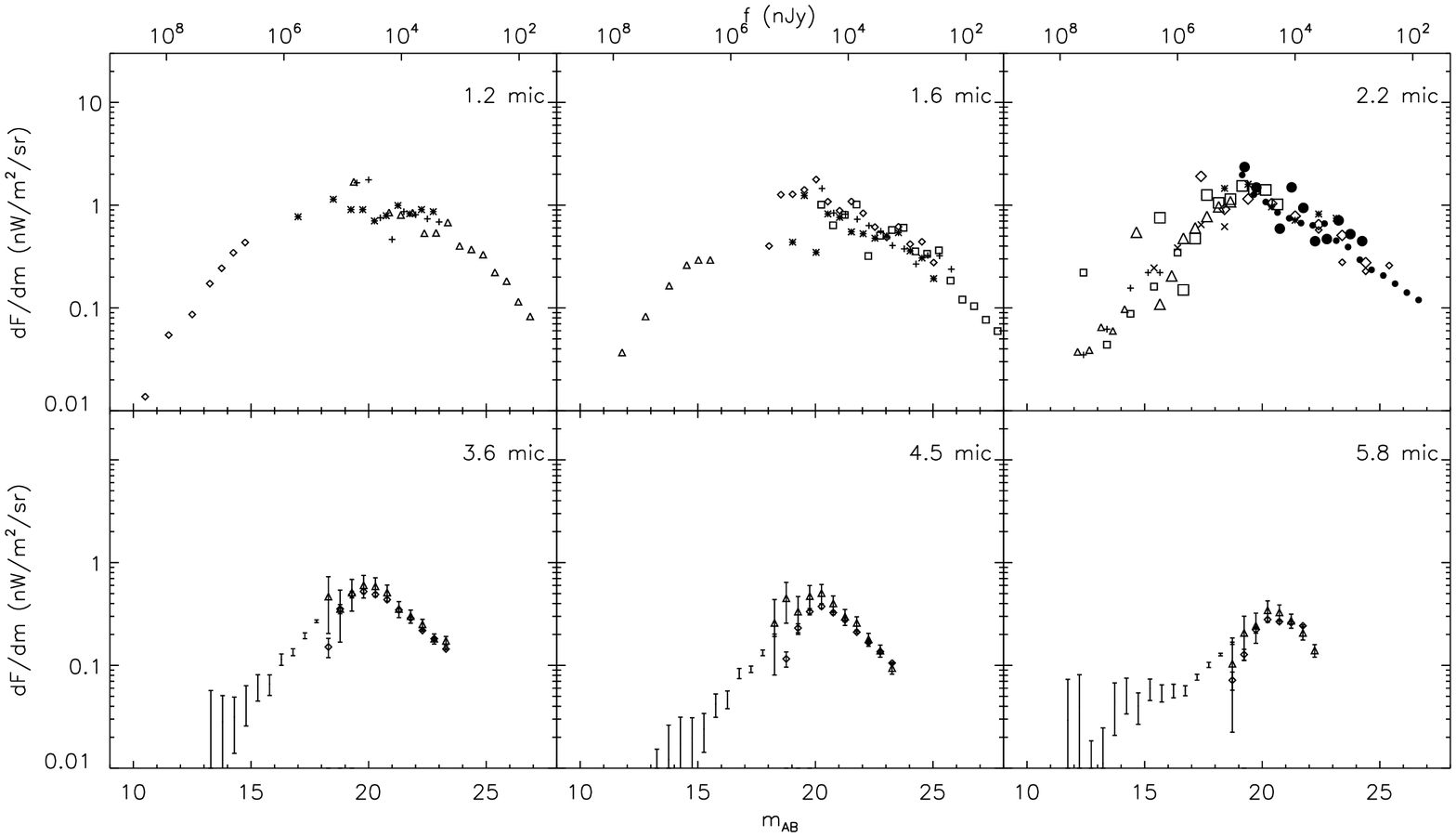}

  \caption{Cumulative flux in nW/m$^2$/sr contributed by galaxies from a narrow
$dm$ magnitude bin for HST (Madau \& Pozzetti 2000 - upper panels)
and Spitzer-based (Fazio et al 2004) counts data. The points are
adopted from Kashlinsky (2005a - see Figs. 15,16).
}\label{fig:cib_counts}
\end{figure}

One can compare the claimed CIB levels to the CIB levels obtained
from integrating the contributions from galaxies seen in faint
galaxy counts surveys. The net fluxes per magnitude interval $dm$
from such counts are given by:
\begin{equation}
\frac{dF}{dm} = f(m) \frac{dN}{dm} \label{eq:counts_cib}
\end{equation}
where $dN/dm$ is the differential counts of galaxies per unit
solid angle in the magnitude bin $dm$ and $f(m)$ is the flux
corresponding to $m$.

Fig. \ref{fig:cib_counts} shows the near-IR CIB flux contributions
per $dm$ from the galaxy populations observed in various deep
surveys. The total NIR fluxes by these "ordinary" galaxies (OG)
from the HST and Spitzer IRAC measurements are shown as filled
circles in Fig. \ref{fig:cib_dc}; these total fluxes saturate at
AB magnitude of about 20 and at IR wavelengths shorter than $\sim
5-8$ \um\ they appear to be lower by a significant factor than the
detected CIB.

The difference between the claimed CIB levels and the total fluxes
from ``ordinary" (i.e. not Pop~III) galaxies are known as the
near-IR background excess (hereafter NIRBE), whose integrated
amount between 1 and 4 \um\ is (Kashlinsky 2005a):
\begin{equation}
F_{\rm NIRBE}(1\!-\!4\mu{\rm m}) = 29\pm 13 \;\; {\rm nW/m^2/sr}
\label{eq:nirbe} \end{equation}

\section{Cosmic infrared background excess and Population III}
\label{sec:pop3-theory}

It was suggested that the NIR CIB excess is produced by massive
Pop III stars at high $z$ ($> 8-10$)
(\cite{santos,salvaterra,ferrara,cooray,kagmm}). Because Pop III
stars, if massive, would radiate at the Eddington limit, where
$L\propto M$, the levels of the total flux produced by them are
largely model-independent (\cite{rees,kagmm}) leading to a robust
prediction for the total bolometric flux from them. For
completeness we reproduce briefly the argument from Kashlinsky et
al (2004): Each star would produce flux $\frac{L}{4\pi d_L^2}$,
where $d_L$ is the luminosity distance. Because for massive stars
$L\propto M$, the total comoving luminosity density from Pop III
is $\int n(L) L dL \propto \Omega_{\rm baryon} f_*
\frac{3H_0^2}{8\pi G}$, where $n(L)$ is their luminosity function
and $f_*$ is the mass fraction of baryons locked in Pop III stars
at any given time. In the flat Universe, the volume per unit solid
angle subtended by cosmic time $dt$ is $dV =c(1+z) d_L^2dt$.
Finally, these stars would radiate at efficiency $\epsilon$
($\simeq 0.007$ for hydrogen burning). This then leads to the
closed expression for the total bolometric flux from these
objects:
\begin{equation}
F_{\rm bol} = \frac{3}{8\pi} \frac{c^5/G}{4\pi R_H^2}
\langle(1+z)^{-1}\rangle \epsilon f_3 \Omega_{\rm baryon} \simeq
4\times 10^7 z_3 \epsilon f_3 \Omega_{\rm baryon}h^2 \frac{\rm
nW}{\rm m^2 sr}
\end{equation}
Here $f_3$ is the mean mass fraction of baryons locked in Pop III
stars and $z_3 \equiv \frac{1}{\langle (1+z)^{-1}\rangle}$ is a
suitably averaged redshift over their era. The total flux is a
product of the maximal luminosity produced by any gravitational
process, $c^5/G$, distributed over the surface of the Hubble
radius, $R_H$=$cH_0^{-1}$, and the fairly understood dimensionless
parameters. The term $c^5/G$ appears because the emissions are
produced by the nuclear burning of stars evolving in gravitational
equilibrium with the (radiation) pressure (or gravitational
accretion as in the case of mini-quasars). From WMAP observations
we adopt $\Omega_{\rm baryon}h^2$=0.044 (\cite{spergel}) and,
because the massive stars are fully convective, their efficiency
is close to that of hydrogen burning (\cite{schaerer,siess}),
$\epsilon$=0.007.

Requiring that the massive Pop III stars are responsible for the
flux given by eq. \ref{eq:nirbe} leads to the fraction of baryons
locked in them of:
\begin{equation}
f_3 = (4.2 \pm 1.9)\times 10^{-3}  z_3 \frac{0.044}{\Omega_{\rm
baryon}h^2}\frac{0.007}{\epsilon} \label{eq:pop3fraction}
\end{equation}
 Within the uncertainty of
eq. \ref{eq:nirbe}, only $\sim 2\%$ of the baryons had to go
through Pop III in order to produce the entire NIRBE. This is not
unreasonable considering that primordial clouds are not subject to
many of the effects inhibiting star formation at the present
epochs, such as magnetic fields, turbulent heating etc. The only
criterion for Pop III formation seems to be that primordial clouds
turning-around out of the ``concordance" $\Lambda$CDM density
field have the virial temperature, $T_{\rm vir}$, that can enable
efficient formation of and cooling by molecular hydrogen
(\cite{abel,bromm}). Assuming spherical collapse of gaussian
fluctuations and the $\Lambda$CDM model from WMAP observations
(\cite{spergel}) the fraction of collapsed haloes at $z$=10 with
$T_{\rm vir}\geq$(400,2000)K is $(2.6,5)\times 10^{-2}$, which is
in good agreement with eq. \ref{eq:pop3fraction}. These stars
would have to be dominated by masses $> 240 M_{\odot}$ to be
consistent with low metallicities observed in Population II
(\cite{heger}).

The fraction given by eq. \ref{eq:pop3fraction} was evaluated by
using the bolometric CIB excess integrating from 1 to 4\um. If
there were significant amounts of CIB excess flux missed outside
that range, the fraction $f_3$ would increase. However, at
wavelengths $<0.1z_3$\um\ the emission from the early times would
be below the Lyman break and would be reprocessed to $\lambda >
1$\um\ (Santos et al 2002). If significant Pop III activity
continued at $z_3<10$, which is unlikely, the rest frame Lyman
break may be redshifted to $<$1\um, but the possible extra CIB
excess from $<$1\um\ will be compensated for by $f_3$ in eq.
\ref{eq:pop3fraction} decreasing with $z_3$. At longer
wavelengths, the CIB excess given by eq. \ref{eq:nirbe} can at
most increase $f_3$ by $\sim 30\%$. Thus the entire NIRBE can be
explained if 2-4 \% of the baryons have been converted into
massive Pop~III stars at $z> 10$ (Madau \& Silk 2005, Kashlinsky
2005b). This fraction of converted baryons scales linearly with
the amplitude of $F_{\rm NIRBE}$.

\section{CIB excess and high energy gamma-ray absorption}
\label{sec:gamma-rays}

An independent limit on the net amount of the isotropic component
of the CIB comes from the measurements of absorption of
high-energy cosmological sources. The latter results from the
two-photon absorption, $\gamma \gamma_{\rm CIB} \rightarrow
e^+e^-$, effective at high gamma-ray energies, $E_{\gamma}
E_{\gamma_{\rm CIB}}> (m_ec^2)^2$. Because the reaction is due to
electromagnetic interaction, its cross section is of order $\sim$
(the electron radius, $e^2/m_ec^2$)$^2$ (it peaks at
$\sim\frac{1}{4}\sigma_T$) which leads to significant absorption
over cosmological distances.

\begin{figure}
 \includegraphics[height=2.75in,width=5.6in]{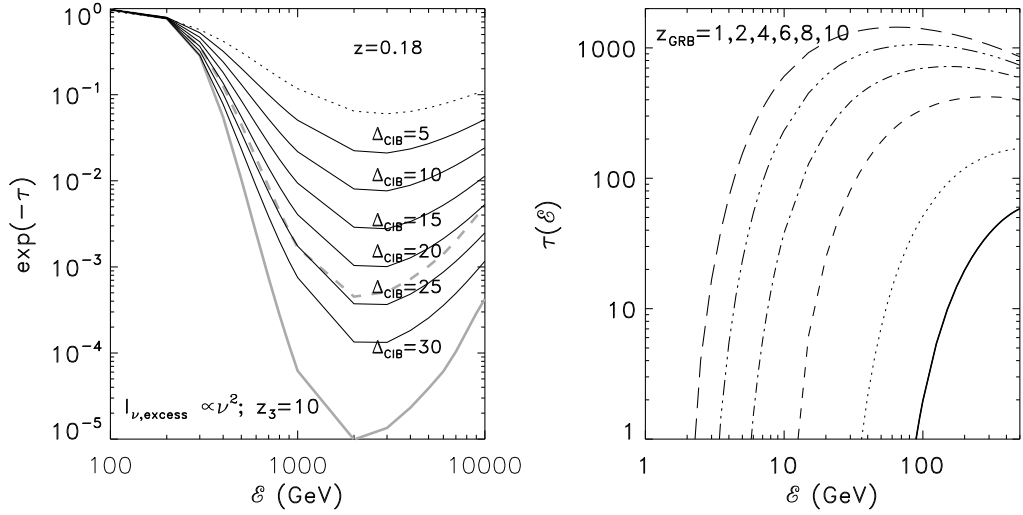}

  \caption{{\it
Left}: Optical depth to photon-photon absorption for a source at
$z$=0.18. The convention for the thick light-shaded lines
corresponds to the inset in Fig.~1. Dotted line assumes only
ordinary galaxies measured in deep counts and that their photons
originated at $z\geq0.18$. Solid lines correspond to the near-IR
CIB excess from the Pop III era assuming it ended at $z_3=10$. The
values of the excess in nW/m$^2$/sr are marked near the lines.
{\it Right}: The net $\tau$ vs the GRB photon energy for the GRB
redshifts shown in the panel. Solid, dotted, short-dashed,
dash-dotted, dash-triple-dotted  and long-dashed lines correspond
to increasing order in $z$.
 }\label{fig:grbs}
\end{figure}

The recent results from two $z\sim 0.18$ blazar spectra
measurements at TeV energies by the HESS collaboration (Aharonian
et al 2006) and similar constraints using slightly lower $z$
blazars (Dwek et al 2005) indicate that the claimed NIRBE would
lead to more attenuation at $\sim 1-2$ TeV than the known blazar
physics would allow. The thick light-shaded lines in the left
panel of Fig. \ref{fig:grbs} show the attenuation due to NIRBE and
galaxy counts fluxes for the CIB going through the central points
of the near-IR CIB measurements (solid thick light-shaded line)
and through the lower edges of the data (dashed thick light-shaded
line). Such high attenuation factors (in excess of $\sim 10^3$) at
$\sim 1.5$ TeV have been argued against by the HESS results. This
would indicate that - barring any changes in blazar physics - the
NIRBE levels may be smaller than given by eq. (\ref{eq:nirbe}).

However, it is important to emphasize that the HESS data still
require significant CIB fluxes from the Pop~III era. Indeed, this
is also shown in Fig.~\ref{fig:grbs} where we computed the
attenuation factors assuming that the NIRBE contribution from
Pop~III scales as $I_\nu \propto \nu^2$ with a Lyman limit cutoff
corresponding to the Pop~III era ending at $z_3$=10 and normalized
to the shown levels of the NIRBE flux, $\Delta_{CIB}$ in
nW/m$^2$/sr. The attenuation due to CIB from ordinary galaxies
alone (filled circles in Fig. \ref{fig:cib_dc}) is shown as dotted
line and is {\it not enough} to account for the attenuation in the
HESS blazars. The attenuation due to CIB levels claimed by the
IRTS and DIRBE measurements is probably too strong. However, the
figure shows that {\it significant amounts of NIRBE are still
allowed and required by the data}. In particular the HESS data
requires the levels of NIRBE due to Pop~III (i.e. with Lyman
cutoff in the CIB at 1 \um) to be below $\sim$15-20 nW/m$^2$/sr
still leaving up to a few percent of the baryons to have gone
through Pop~III.

This situation is expected to be resolved with the data from the
upcoming GLAST mission which should measure spectra of high-$z$
gamma-ray bursts (GRBs) and blazars out to 300 GeV (Kashlinsky
2005b). If Pop III at early epochs produced even a fraction of the
claimed NIR CIB excess, they would provide a source of abundant
photons at high $z$. The present-day value of $I_\nu$=1 MJy/sr
corresponds to the comoving number density of photons per
logarithmic energy interval, $d\ln E$, of
$\frac{4\pi}{c}\frac{I_\nu}{h_{\rm Planck}}$=0.6 cm$^{-3}$ and, if
these photons come from high $z$, their number density would
increase $\propto (1+z)^3$ at early times. These photons with the
present-day energies, $E$, would also have had higher energies in
the past: $E^\prime = E(1+z)\simeq (0.1-0.3)(1+z)$eV. They would
thus provide abundance of absorbers for sources of sufficiently
energetic photons at high redshifts. Regardless of the precise
amount of the NIR CIB from them, Pop~III objects likely left
enough photons to provide a large optical depth for high-energy
photons from distant GRBs. The right panel in Fig. \ref{fig:grbs}
comes from  Kashlinsky (2005b) and shows the net optical depth
(normalized to eq. \ref{eq:nirbe}) at high $z$. It shows that even
if the NIRBE levels from Pop~III were significantly smaller than
in eq. \ref{eq:nirbe} there should still be almost complete
damping in the spectra of high-$z$ gamma ray sources at energies
$< 260 (1+z)^{-2}$ GeV. Such damping should provide an unambiguous
feature of the Pop~III era and GLAST observations expected during
the coming years would provide important information on the
emissions from the Pop~III era.

\section{CIB fluctuations from deep Spitzer images and early populations}
\label{sec:spitzer_kamm}

Population III, if massive, should also have left significant CIB
fluctuations providing a unique signature of their existence
(\cite{cooray,kagmm}). The reasons strong CIB fluctuations are as
follows:

$\bullet$ If massive, each unit of mass in Pop III stars would
emit several order of magnitude more luminosity than the
present-day stellar populations.

$\bullet$ Pop III era is expected to have spanned a relatively
short cosmic time ($<$ 1 Gyr) leading to larger relative CIB
fluctuations.

$\bullet$ Pop III stars are expected to have formed out of rare
high-$\sigma$ peaks of the density field which would amplify their
clustering.

We (Kashlinsky, Arendt, Mather, Moseley 2005) have attempted to
measure these CIB fluctuations in deep Spitzer/IRAC data at 3.6 to
8 \um. The results of these measurements and their interpretation
are discussed below. New and deeper measurements (Kashlinsky,
Arendt, Mather, Moseley 2006, in preparation) support our earlier
results.

\subsection{Observational results}

The main data used in Kashlinsky et al (2005; hereafter KAMM) came
from the IRAC guaranteed-time-observations observations with
$\sim$10 hour integration of a field at high Galactic latitude.
Additionally, the available data with shallower observations for 2
auxiliary fields were analyzed in order to test for isotropy of
any cosmological signal. The datasets were assembled out of the
individual frames using the least calibration method which has
advantages over the standard pipeline calibration of the data in
that the derived detector gain and offsets match the detector at
the time of the observation, rather than at the time of the
calibration observations.

For the final analysis we selected a subfield of $\simeq 5^\prime
\times 10^\prime$ with a fairly homogeneous coverage. Individual
sources have been clipped iteratively. The images were left with
$> 75\%$ pixels for robust computation of the diffuse flux Fourier
transforms. CIB fluctuations from Pop~III at high $z$ should be
independent of the clipping threshold, so the maps were also
clipped progressively deeper to verify that our results are
threshold-independent. As more pixels are removed, it becomes
impossible to evaluate robust Fourier transforms, and then the
diffuse flux correlation function was calculated. In that analysis
we (\cite{spitzer}) detected fluctuations significantly exceeding
the instrument noise.

\begin{figure}
 \includegraphics[height=3in,width=5.6in]{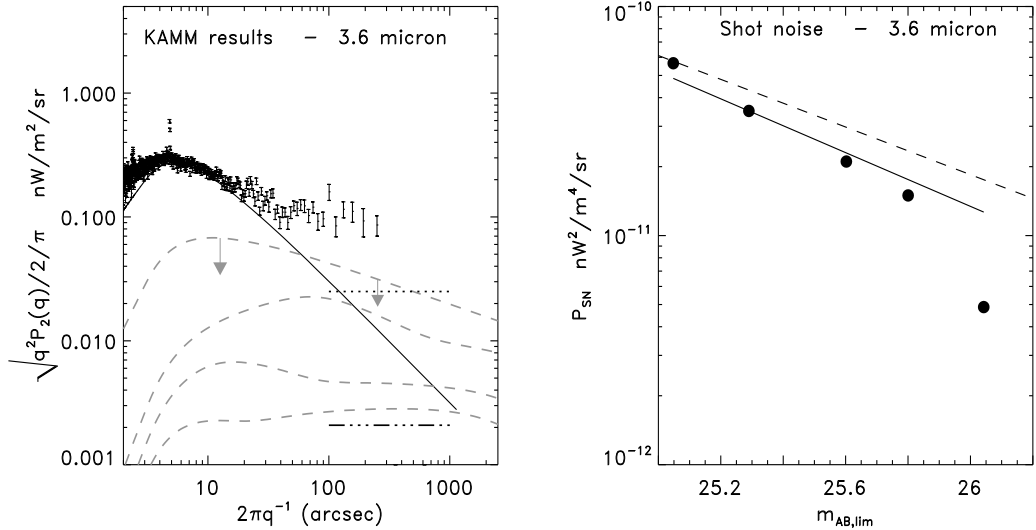}
 \caption{{\it
Left}: the CIB fluctuations spectrum measured by KAMM at 3.6 \um\
are shown with 1-$\sigma$ error bars. Solid line shows the
contribution from the shot noise from remaining ordinary galaxies.
The upper limit on the zodiacal light fluctuations are shown with
the dashed-triple-dotted line and on Galactic cirrus contribution
with the dotted line. The upper limit on the contribution from
ordinary galaxies is shown with the upper dashed line marked with
down-pointing arrows. The lower dashed lines show more realistic
contributions from ordinary galaxy populations as described in
KAMM. {\it Right}: The amplitude of the shot noise power spectrum
from remaining ordinary galaxies shown vs the limiting magnitude
of galaxy removal by KAMM. Solid line is from the KAMM fit to the
Spitzer counts (Fazio et al 2004); dashed line corresponds to
counts analysis of Savage \& Oliver (2005)}\label{fig:kamm}
\end{figure}

Fig. \ref{fig:kamm} shows the CIB fluctuation, after the power of
the instrument noise has been subtracted, at 3.6 $\mu$m. The
excess fluctuation on arcminute scales in the 3.6 $\mu$m channel
is $\sim 0.1$ nW/m$^2$/sr; KAMM measure a similar amplitude in the
longer IRAC bands indicating that the energy spectrum of the
arcminute scale fluctuations is flat to slowly rising with
increasing wavelength at least over the IRAC range of wavelengths.
The detected signal is significantly higher than the instrument
noise and the various systematics effects cannot account for it.
There was a statistically significant correlation between the
channels for the region of overlap suggesting the presence of the
same cosmic signal in all channels. The correlation function at
deeper clipping cuts, when too few pixels were left for Fourier
analysis, remains roughly the same and is consistent with the
power spectrum numbers.

Based on numerous tests KAMM show that the signal comes from the
sky. Below are the possible sources of the fluctuations; they
include origin in the spacecraft, the Solar system, the Galaxy and
the (cosmological) sources outside the Galaxy:

1) Instrument noise was estimated in KAMM, who show that it is too
low and has different pattern than the detected signal. Also, the
cross-correlation between the channels for the overlap region of
the data rules out significant instrumental contribution.

2) Residual wings of removed sources also appear unlikely to have
contributed significantly to the detected signal. KAMM have done
extensive analysis and many tests to rule this out. E.g. the
results remain the same for various clipping parameters, and for
the various masks of the clipped-out sources as shown in their
Supplementary Information.

3) Zodiacal light fluctuations were estimated by subtracting
another data observed at two epochs separated by $\sim$ 6 months.
These are shown in Fig. \ref{fig:kamm} and are very small: the
amplitude at 8 mic is $<0.1$ nW/m$^2$/sr and assuming normal zodi
spectrum the contribution to fluctuations would be totally
negligible at shorter wavelengths.

4) Galactic cirrus is significant at channel 4 (8 \um), which may
in fact be dominated by the cirrus component (diffuse flux
fluctuation $\sim 0.2$ nW/m$^2$/sr), but given the energy spectrum
of cirrus emission the other channels should have negligible
cirrus. However, the map at Channel 4 correlates statistically
significantly with the shorter wavelengths suggesting the presence
of the same cosmic signal at 8 \um\ as well.

5) Extragalactic sources thus seem the remaining logical
explanation of the results. They include:

     5.1)  Ordinary galaxies with "normal" stellar populations

     5.2)  Population III with $M/L << (M/L)_\odot$ and located at
     high $z$.

We discuss the constraints the data impose on the two cosmological
components. More discussion is given with new measurements in
Kashlinsky, Arendt, Mather \& Moseley (2006, in preparation).

\subsection{Interpretation}
\label{sec:interpretation}

Extragalactic contributions to the CIB fluctuations come in two
flavours: 1) the shot noise, and 2) due to clustering out of the
primordial density fields which these sources trace. The
shot-noise contribution, when evaluated directly from galaxy
counts, gives a good fit to the fluctuations on smaller scales as
shown in Fig. \ref{fig:kamm}. Ordinary galaxies have been
eliminated from the maps down to very faint flux levels ($\sim
0.3\mu$Jy in Channel 1). The remaining ordinary galaxies'
contribution to the CIB mean levels is small ($<0.1-0.2$
nW/m$^2$/sr) and its fluctuation levels are shown in Fig.
\ref{fig:kamm}. On the other hand, the amplitude, power spectrum
and the spectral energy dependence of the $>$ arcminute-scale
fluctuation can be explained by emissions from Pop~III.

\begin{figure}
 \includegraphics[height=2.75in,width=5.in]{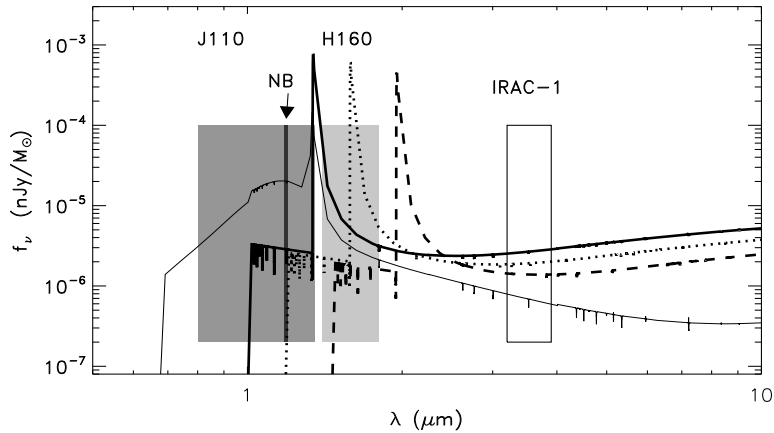}
  \caption{SED from Pop~III system is shown for $z$=10 (thick
solid line), 12 (dotted) and 15 (dashed). The lines are drawn from
Santos et al (2002) for the case when processing of the radiation
takes place in the gas inside the nebula. Thin solid line shows
the SED from Santos et al for $z$=10 when the nebula is assumed
transparent and the processing takes place inside the IGM. The J,
NB, H and IRAC-3.6\um\ filters are shown. When the emission is
reprocessed entirely by the IGM, the emission could extend below
the Lyman limit at the rest-frame of the emitters, in which case
many sources could have existed even at $z$=10, but escaped
detection via the $J-NB>0.3$ criterion of ZEN (Willis \& Courbin
2005).}\label{fig:pop3sed}
\end{figure}

Fig. \ref{fig:pop3sed} shows the predicted flux due to spectral
energy distribution (SED) of a Population III stellar system at
$z=12-15$ and the region probed by the IRAC data. Although the
Spitzer instruments cannot probe the peak of the emissions (due to
Lyman emission), at the IRAC wavelengths we are still probing a
region with substantial emissions by the Population III sources.

Any interpretation of the KAMM results must reproduce three major
aspects:

$\bullet$ The sources in the KAMM data were removed to a certain
(faint) flux limit, so the CIB fluctuations arise in populations
with magnitudes fainter than the corresponding magnitude limit,
$m_{\rm lim}$.

$\bullet$ These sources must reproduce the excess CIB fluctuations
by KAMM on scales $> 0.5^\prime$.

$\bullet$ Lastly, the populations fainter than the above magnitude
limit must account not only for the correlated part of the CIB,
but - equally importantly - they must reproduce the (low)
shot-noise component of the KAMM signal, which dominates the power
at $<$0.5$^\prime$.

Below we briefly discussion the constraints on the populations
contributing to the KAMM signal in the above order:

1) {\bf Magnitude limits}. The nominal limit above which sources
have been removed in the KAMM analysis is $m_{\rm AB}$=25.3 at 3.6
um and this by itself implies that the detected CIB fluctuations
arise from fainter systems. At this magnitude one is already at
the confusion limit for the IRAC beam at 3.6 \um\, so fainter
galaxies can be excised only by removing significantly more area
in the map. The new mask then prevents computing reliably the
power spectrum of the diffuse flux, which is why for deeper source
removal KAMM presented its Fourier transform, the correlation
function $C(\theta)$, in the Supplementary Information (SI). What
is important in this context is that the $C(\theta)$, shown in
Fig. SI-4 of the KAMM SI, does not change in shape and amplitude
with additional clipping. At 3.6\um\ the extra clipping goes to
the limit of $\simeq$140nJy for the faintest removed sources when
KAMM stopped as only 6\% of the map pixels remained. Thus the KAMM
signal comes from sources fainter than $m_{\rm AB}
> 26.1$. For reference, this magnitude limit corresponds to
$6 \times 10^8 h^{-2} L_\odot$ emitted at 6000 \AA\ at $z$=5. A
significant fraction of galaxies were thus removed from the data
by KAMM even at $z\geq 5$ and the detected CIB fluctuations must
be explained by still fainter and more distant systems.

 2) {\bf Clustering component} from the remaining sources is given
 by the Limber equation which we write in the form (Kashlinsky 2005a):
\begin{equation}
\left[\frac{q^2 P_2(q)}{2\pi}\right]^{1/2} = F_{\rm CIB}
\Delta(q\langle d_A^{-1}\rangle) \; \;\; ; \; \Delta^2(k) \equiv
\frac{1}{2\pi} \frac{k^2P_3(k)}{c\Delta t} \label{eq:limber}
\end{equation}
where $P_3(k)$ is the 3-D power spectrum of the sources'
clustering, $d_A$ is the comoving angular diameter distance and
$\Delta t$ is the cosmic time interval spanned by the sources.
Fig. \ref{fig:kamm} shows that the clustering strength at $\geq
1^\prime$ requires $\delta F_{\rm CIB}\sim 0.1$ nW/m$^2$/sr. The
angle of $1^\prime$ in the concordance cosmology subtends comoving
scales of 2.2-3 Mpc at 5$\leq z \leq$20. For $\Lambda$CDM density
fields with reasonable biasing one can reach relative
arcminute-scale fluctuations of $\leq$5-10\% meaning that the net
CIB from sources contributing to the KAMM signal at 3.6 \um\ is
$>$1-2 nW/m$^2$/sr.

3) {\bf Shot noise constraints}. The amplitude of the shot-noise
power gives a particularly strong indication of the epochs of the
sources contributing to the KAMM signal. This can be seen from the
expressions for the shot noise (Kashlinsky 2005a):
\begin{equation}
P_{\rm SN} = \int_{m\!>\!m_{\rm lim}}f(m) dF(m) \equiv f(\bar{m})
F_{\rm tot}(m\!>\!m_{\rm lim})\label{eq:shotnoise}
\end{equation}
Fig. \ref{fig:kamm} shows the shot noise amplitude evaluated at
3.6 \um\ for the KAMM data as function of the source removal
threshold. Because the shot-noise sets important limits on the
contributions to the CIB fluctuations from remaining ordinary
galaxies, we have evaluated $P_{\rm SN}$ for the data at deeper
clipping limits as follows: the maps analyzed and shown in KAMM's
SI were clipped to progressively lower limits until $\simeq 6\%$
of the map remained. At deeper clipping thresholds, where one
cannot compute Fourier transforms reliably, we evaluated $P_{SN}$
assuming it is proportional to the variance of the map,
$\sigma_F^2$, minus that of the noise, $\sigma_n^2$, estimated by
KAMM from the difference maps. (The noise, as expected, does not
vary with the regions removed by deeper clipping and at 3.6 \um\
remains $\sigma_n \simeq 0.5$ nW/m$^2$/sr.) The last point may be
an under-estimate as the precise value of the shot noise there may
be influenced by not sufficiently well determined (for that
purpose) instrument noise levels.

From Fig. \ref{fig:kamm} we conservatively adopt the fiducial
value of $P_{\rm SN}=10^{-11}$ nW$^2$/m$^4$/sr as the {\it upper}
limit to the shot noise levels of the sources contributing to the
measured CIB fluctuations. Above it was shown that the sources
contributing to the fluctuations must have CIB flux greater than a
few nW/m$^2$/sr and combining this with the values for $P_{\rm
SN}$ shown leads via eq. \ref{eq:shotnoise} to these sources
having typical magnitudes $m_{\rm AB} > 29-30$ or individual
fluxes below a few nJy. {\it Such faint sources are expected to
lie at $z> 10$, the epoch of the first stars.}

\section{Resolving the sources of the cosmic infrared
background} \label{sec:confusion}

The CIB arises from emissions by discrete cosmological sources is
therefore not really diffuse. It can and, of course, should be
resolved eventually with faint/deep surveys which simultaneously
have sufficiently fine angular resolution.

Salvaterra \& Ferrara (2006) have recently raised an important
question of whether the existence of the Pop~III era is in
conflict with the recent data on J band dropouts (Bouwens et al
2005, Willis \& Corbin 2005). The surveys are strongly incomplete
at $H> 26$, but, at face value, the data indicate a paucity of
$J-H>1.8$ dropouts around $H_{\rm AB}\simeq 28$ of only at most
$\sim$ a few per arcmin$^2$ (Bouwens et al 2005) and a similarly
low upper limit on the dropouts at $J-NB\geq 0.3$ in the ZEN
survey of Willis \& Corbin (2005). Is there necessarily a conflict
or do these data tell us something significant about the Pop III
era physics? Such theoretical computations of the counts of the
Pop III sources are - for now - necessarily based on the following
assumptions listed here in the decreasing order of the
subjectively perceived likelihood: 1) the distribution of small
scale fluctuations of the primordial density field is Gaussian; 2)
the small scale power is that of the concordance $\Lambda$CDM
model; 3) the mass function of the first objects is described well
by a Press-Schechter type prescription although the effective
index of the $\Lambda$CDM power spectrum at these scales
$n\rightarrow -3$ and pressure effects may be important at such
low masses (cf. Springel et al 2005); 4) that the redshift, $z_3$,
out to which Pop~III era extended is $z_3< 9$, 5) the SED of
Pop~III systems is cut off at the Lyman-$\alpha$ frequency of the
source epoch and the cutoff is not produced by the IGM at lower
$z$ (cf. Santos et al 2002), and most crucially 6) that the
efficiency of Pop~III formation inside the parental halos is the
same for all masses and epochs. Dropping any of the assumptions
would lead to significant changes in the predicted number counts.
The last assumption is particularly critical, because of the ease
of both the destruction of $H_2$ molecules by Lyman-Werner band
photons (e.g., Haiman et al. 1997) and their creation following
the ionization of the surrounding nebula (Ferrara 1998; see
reviews by Barkana \& Loeb 2001; Bromm \& Larson 2004).

Fig. \ref{fig:pop3sed} marks the filters around 1.1 and 1.6 \um\
used by Bouwens et al (2005) and Willis \& Corbin (2005) vs an
example of the emission template from Pop~III stars at
$z$=10,12,15. The figure shows that both surveys probe a very
narrow range of redshifts: $z$=12 by Bouwens et al and $z$=10 by
ZEN. Indeed, only for $z$=12 are sufficiently massive Pop~III
systems (the mass in Pop III stars $M_*> 10^6M_\odot$) likely to
produce $H\sim$27-28 and satisfy $J-H>$1.8. (The constraints from
the ZEN survey are weaker). Thus the counts problem - even with
the above assumptions - provides information only for these narrow
($\Delta z<\sim 1$) epochs. The discussion in Sec.
\ref{sec:interpretation} shows that Pop~III systems have the mean
flux of less than $\sim 5-10$ nJy at 3.6 \um; such populations
cannot be detected in the above surveys. Why Pop~III form in such
small systems  is an interesting question which is likely related
to the variation of the efficiency of Pop~III formation with epoch
and mass.

In order to detect the faint sources responsible for the CIB
fluctuations with fluxes below a few nJy, as discussed in Sec.
\ref{sec:interpretation}, embedded in the underlying sea of
emissions, their individual flux must exceed the confusion limit
usually taken to be $\alpha\geq 5$ times the flux dispersion
produced by these emissions (Condon 1974). If these sources do not
have the necessary flux levels they will be drowned in the
confusion noise and will not be individually identifiable in
galaxy surveys. Of course, this is precisely where CIB studies
would take off. From observations one knows that confusion levels
are not reached in J band until $m_{\rm AB} \sim 28$ (e.g. Gardner
\& Satyapal 2002). If such sources were to contribute to the CIB
required by KAMM data, at 3.6 \um\ they had to have the average
surface density of
\begin{equation}
\bar{n} \sim F_{\rm CIB}^2/P_{\rm SN} \sim 5 \; {\rm arcsec}^{-2}
\; \left(\frac{F_{\rm CIB}}{2\;{\rm nWm^{-2}sr^{-1}}}\right)^2
\left(\frac{P_{\rm SN}}{10^{-11}\; {\rm
nW^2m^{-4}sr^{-1}}}\right)^{-1} \label{eq:confusion_jwst}
\end{equation}
In order to avoid the confusion limit and resolve these sources
individually at, say, 5-sigma level ($\alpha=5$) one would need a
beam of the area
\begin{equation}
\omega_{\rm beam} \leq \alpha^{-2}/\bar{n} \sim 5\times
10^{-3}\left(\frac{F_{\rm CIB}}{2\;{\rm
nWm^{-2}sr^{-1}}}\right)^{-2}{\rm arcsec}^2 \label{eq:beam}
\end{equation}
 or of circular radius below
$\sim$0.04 $(F_{\rm CIB}/2{\rm nWm^{-2}sr^{-1}})^{-1}$arcsec. This
is clearly not in the realm of the currently operated instruments,
but the {\it JWST} could be able to resolve these objects given
its sensitivity and resolution.

\section{Conclusions}
\label{sec:conclusions}

Both the CIB and Population III studies are rapidly evolving into
observation-rooted, if not yet high-precision, disciplines. The
current situation is therefore very dynamic, but one can summarize
the up-to-date constraints on the Pop~III era that emerge from the
recent CIB measurements as follows:

$\bullet$ The measurements of the mean levels of the CIB indicate
that about up to a few \% of baryons may have gone through
Pop~III. The new measurements of the $z\sim 0.1-0.2$ blazar
spectra at $\sim$ a few TeV energies indicate that the net levels
of the CIB may be not as high as the DIRBE- and IRTS-based
analyses indicate. Nevertheless, these data show an evidence of
absorption in excess of what can be provided by the observed
galaxy populations indicating additional fluxes from fainter (and
likely more distant) sources. If real, these near-IR CIB fluxes
should be detectable from the future GLAST-based observations of
the high $z$ gamma-ray bursts.

$\bullet$ The recent measurements of the CIB fluctuations allow to
probe the emissions from high $z$ sources more directly. They
indicate that the sources have fluxes below $\sim 10$ nJy at
3.6\um, or $m_{AB}>28-29$. Such sources are likely to be located
at $z> 10$, the era associated with the first stars. These sources
must have provided CIB levels of at least a few nW/m$^2$/sr at 3.6
to 5 \um\ and their clustering properties have to fit the measured
CIB fluctuations of $\sim 0.1$ nW/m$^2$/sr at arcmin scales as
well as have the SED that gives flat to slowly rising CIB
fluctuations with increasing wavelength.

$\bullet$ The above constraint on the typical flux (a few nJy) of
these sources suggests that they have less than $\sim 10^5M_\odot$
in stars inside each of the haloes.  This likely requires a
halo-dependent efficiency if the parental haloes form out of the
$\Lambda$CDM density field prior to $z\sim 10$.

$\bullet$ The sources contributing to the CIB fluctuations at the
Spitzer-IRAC bands should be resolvable in surveys with
sufficiently low noise (to measure the low fluxes) and high
angular resolution (to overcome cosmological confusion). At 3.6
\um\ this can probably be done with the future James Webb Space
Telescope.

{\bf Acknowledgements}. I warmly thank my collaborators on the CIB
fluctuations project, Rick Arendt, John Mather and Harvey Moseley,
for many discussions and contributions. This work was supported by
NSF and NASA/Spitzer grants.

\end{document}